\documentclass[
preprint,
 amsmath,amssymb,
 aps,
prb,
longbibliography
]{revtex4-1}

\usepackage{graphicx}
\usepackage{dcolumn}
\usepackage{bm}
\usepackage{color}
\usepackage{braket}
\usepackage{multirow}
\usepackage[colorlinks,linkcolor=blue, urlcolor=blue,citecolor=blue,bookmarks=fales]{hyperref}

\begin{document}

\title{Searching for Majorana quasiparticles at vortex cores in iron-based superconductors}

\author{Tadashi Machida$^{1,2}$ and Tetsuo Hanaguri$^{1}$}

\affiliation{$^{1}$RIKEN Center for Emergent Matter Science, Wako, Saitama 351-0198, Japan \email{tadashi.machida@riken.jp}}
\affiliation{$^{2}$Precursory Research for Embryonic Science and Technology (PRESTO), Japan Science and Technology Agency (JST), Tokyo 102-0076, Japan}

\begin{abstract}
    The unambiguous detection of the Majorana zero mode (MZM), which is essential for future topological quantum computing, has been a challenge in recent condensed matter experiments. The MZM is expected to emerge at the vortex core of topological superconductors as a zero-energy vortex bound state (ZVBS), amenable to detection using scanning tunneling microscopy/spectroscopy (STM/STS). 
    However, the typical energy resolution of STM/STS has made it challenging to distinguish the MZM from the low-lying trivial vortex bound states. Here, we review the recent high-energy-resolution STM/STS experiments on the vortex cores of Fe(Se,Te), where the MZM is expected to emerge, and the energy of the lowest trivial bound states is reasonably high. 
    Tunneling spectra taken at the vortex cores exhibit a ZVBS well below any possible trivial state, suggesting its MZM origin. However, it should be noted that ZVBS is a necessary but not sufficient condition for the MZM; a qualitative feature unique to the MZM needs to be explored. 
    We discuss the current status and issues in the pursuit of such Majorananess, namely the level sequence of the vortex bound states and the conductance plateau of the ZVBS. We also argue for future experiments to confirm the Majorananess, such as the detection of the doubling of the shot noise intensity and spin polarization of the MZM.\end{abstract}
\maketitle

\section{Introduction}

In condensed matter systems, the Majorana fermion emerges as a zero-energy quasiparticle comprising an equal superposition of an electron and a hole, which is known as Majorana zero mode (MZM). 
Theoretically, it can be realized at a boundary of a topological superconductor as a consequence of the topological bulk-boundary correspondence~\cite{Kitaev_PU_2001,Sato_PL_2003,Nayak_RMP_2008,Liang_RMP_2011,Alicea_RPP_2012,Beenakker_AR_2013,Elliott_RMP_2015,Sato_JPSJ_2016,Sato_RPP_2017,Sau_NRP_2020,Beenakker_SPP_2020}.
The MZM is topologically protected and obeys the non-Abelian statistics, bringing about a potential capability for fault-tolerant quantum computing and stimulating an eager search for its realization and detection~\cite{Nayak_RMP_2008,Liang_RMP_2011,Alicea_RPP_2012,Beenakker_AR_2013,Sato_JPSJ_2016,Sato_RPP_2017,Sau_NRP_2020,Beenakker_SPP_2020,KITAEV_AnPhys_2003,Aasen_PRX_2016}.

A vortex core, which is introduced by an applied magnetic field to a superconductor, can serve as such a boundary accommodating the MZM when the host superconductor is topologically nontrivial~\cite{Volovik_JETP_1999,Read_PRB_2000}. 
In general, at the vortex core, the superconducting gap $\Delta$ locally vanishes, resulting in the formation of vortex bound states. 
In an ordinary $s$-wave superconductor, the energies of these bound states, known as the Caroli-de Gennes-Matricon (CdGM) states~\cite{CAROLI1964307}, are represented by the half-odd-integer level sequence $E~=~\mu \Delta^{2}/\epsilon_{\rm{F}}$ ($\mu~=~\pm1/2, \pm3/2, \pm5/2,\cdots$)($\epsilon_{\rm{F}}$ : Fermi energy) [Fig. 1(a)]. 
It should be noted that there is no state at exactly zero energy.
In contrast, for a chiral $p$-wave superconductor, which is a representative of the topological superconductor, the vortex bound states follow the integer level sequence ($\mu~=~\pm0, \pm1, \pm2,\cdots$), because of the additional angular momentum of the superconducting order parameter. 
Hence, an exactly zero energy state appears for $\mu~=~0$~\cite{Volovik_JETP_1999} [Fig. 1(b)].
This $\mu~=~0$ state is nothing but the MZM.

\begin{figure}[h]
    \centering\includegraphics[width=10cm]{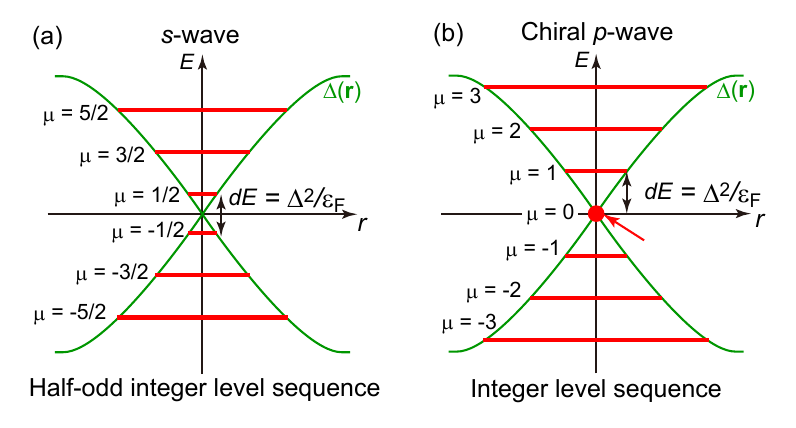}
    \caption{(a, b) Schematic illustrations of energy diagram of vortex bound states for an $s$-wave and a chiral $p$-wave superconductor, respectively. Red and green lines represent the vortex bound states and spatial variation of the superconducting gap, respectively.}
    \label{fig_sim}
\end{figure}

Obviously, a key signature of the putative vortex MZM is the zero-energy vortex bound state (ZVBS), which can be detected by STM/STS as a peak in the local density-of-states spectrum at the vortex core.
However, this experiment is challenging because, for typical superconductors with $\Delta~\sim~1$~meV and $\epsilon_{\rm{F}}~\sim~1$~eV, the energy spacing between the bound states $\delta E~\sim~\Delta^{2}/\epsilon_{\rm{F}}$ is as small as $1~\mu$eV, while the currently available energy resolution of STM/STS is limited to be tens of $\mu$eV at the best.
In such a situation, many vortex bound states crowd inside the superconducting gap, forming a broad peak at zero energy in the tunneling spectrum.
This peak does not represent the ZVBS associated with the MZM but is a resolution-smeared bundle of multiple CdGM states, which should appear irrespective of the topological nature of superconductivity.
Therefore, the experimental identification of the MZM demands the following conditions: 
(i) the superconductor should be the chiral $p$-wave superconductor,
(ii) the superconductor should possess large $\Delta$ and/or small $\epsilon_{\rm{F}}$ to have the large enough energy spacing $\delta E$, and
(iii) the STM/STS should have a high enough energy resolution to resolve $\delta E$.

The recent discoveries of possible chiral $p$-wave superconductivity on the surface of some iron-based superconductors with large $\Delta$ and small $\epsilon_{\rm{F}}$~\cite{Hao_PRX_2014,Wang_PRB_2015,Wu_PRB_2016,Xu_PRL_2016,Hao_NSR_2018,Zhang_Science_2018,Liu_PRX_2018,Liu_NC_2020,Zhang_NP_2019} as well as the improvement of the energy resolution of STM/STS~\cite{Machida_RSI_2018} have brought hope to resolve the discrete bound states and to capture the ZVBS indicative of the MZM.
In this article, we review recent STM/STS investigations of the ZVBS in the iron-based superconductor Fe(Se,Te).
In section 2, we review the basic properties of Fe(Se,Te) focusing on its topological characters and large $\delta E$.
In section 3, we review the current experimental understanding of the ZVBS in Fe(Se,Te) and its relevance to the MZM.
In section 4, we discuss experiments toward unambiguous identification of the MZM.
A summary is given in section 5.

\section{$\mathrm{Fe(Se,Te)}$ as the platform of the MZM}

The realization of chiral $p$-wave superconductivity is the first step toward realizing the vortex MZM.
Although there are some candidate materials, there is no well-established bulk chiral $p$-wave superconductor in nature. 
Therefore, a synthetic platform that hosts effective chiral $p$-wave superconductivity is indispensable.
Among various ways so far proposed\cite{Kitaev_PU_2001,Sato_PL_2003}, the so-called Fu-Kane proposal is regarded as one of the most promising routes~\cite{Fu_PRL_2008}.
Fu and Kane considered a heterostructure consisting of a topological insulator and an $s$-wave superconductor and showed that proximity-induced superconductivity in the spin-polarized Dirac surface state of the topological insulator is effectively chiral $p$-wave type.
Thus the MZM should emerge in the vortex core.
Based on this idea, various heterostructures (\textit{e.g.}, Bi$_2$Te$_3$ thin films on NbSe$_2$) have been examined by STM/STS~\cite{Xu_PRL_2014,Xu_PRL_2015}.
A broad peak has been observed at zero energy in the tunneling spectra.
However, its relevance to the MZM is ambiguous because the energy resolution of the experiments is not enough to identify each bound state separately.
Moreover, in the heterostructure systems, there is an unavoidable problem that the interface is buried under either a topological insulator or a superconductor, which makes the direct detection of the MZM difficult using surface-sensitive STM/STS.

This problem can be solved in the superconducting topological semimetal having the spin-polarized topological surface state by itself~\cite{Hao_NSR_2018,Bian_NC_2016,Guan_SAdv_2016,Sakano_NC_2015,Iwaya_NC_2017}.
In such a system, the proximity effect from bulk superconductivity can open the superconducting gap at the spin-polarized surface states.
Therefore, the effective chiral $p$-wave superconductivity is expected at the surface, allowing us to access the MZM using STM directly. 
In short, this system can be regarded as a natural hybrid system consisting of the bulk trivial superconductor and the topological surface state and thus is called as a \textit{connate} topological superconductor~\cite{Hao_NSR_2018}.

Although there are some candidates of such connate topological superconductors (\textit{e.g.}, $\beta$-PdBi$_2$ and PbTaSe$_2$), most of them suffer from the issue of small $\delta E$, preventing the detection of the ZVBS isolated from the trivial bound states at finite energies~\cite{Guan_SAdv_2016,Iwaya_NC_2017}.
A connate topological superconductor with large enough $\delta E$ was highly anticipated for the detection of the vortex MZM by using STM.

The iron-based superconductors have been actively studied from the viewpoints of both unconventional and topological superconductivity~\cite{Si_NRM_2016,Shibauchi_JPSJ_2020,Fernandes_Nature_2022}.
Among them, Fe(Se,Te) is a gift from nature that satisfies conditions of both connate topological superconductivity and large enough $\delta E$.
Fe(Se,Te) has a simple crystal structure, as shown in Fig. 2(a) and (b).
Low-energy electronic states are governed by the iron 3$d$ orbitals ($d_{xz}, d_{yz}, d_{xy}$) that form hole and electron bands around $\Gamma$ and $M$ points, respectively [Fig. 2(c) and (d)].
A strong correlation of $d$-electrons causes significant band renormalization around the Fermi level, resulting in a very small $\epsilon_{\rm{F}}$~\cite{Tamai_PRL_2010,Rinott_SAdv_2017}.
There is a consensus that the conventional electron-phonon mechanism can not explain the observed high transition temperature $T_{\rm{c}}$, and the microscopic mechanisms based on the spin fluctuations~\cite{Mazin_PRL_2008,Kuroki_PRL_2008,Seo_PRL_2008,Chen_PRL_2009,Maier_PRB_2011,Khodas_PRL_2012,Agterberg_PRL_2017,Paglione_NP_2010,Hirschfeld_RPP_2011} and the orbital fluctuations~\cite{Kontani_PRL_2010,Yanagi_PRB_2010} have been proposed.
Although the actual superconducting mechanism of Fe(Se,Te) remains elusive, the superconducting gap is fully open, suggesting that the Fu-Kane proposal should work if the spin-polarized topological surface state is formed~\cite{Hanaguri_Science_2010,Liu_PRL_2019,Ding_EPL_2008,Richard_JPhys_2015}.
\begin{figure}[h]
    \centering\includegraphics[width=14cm]{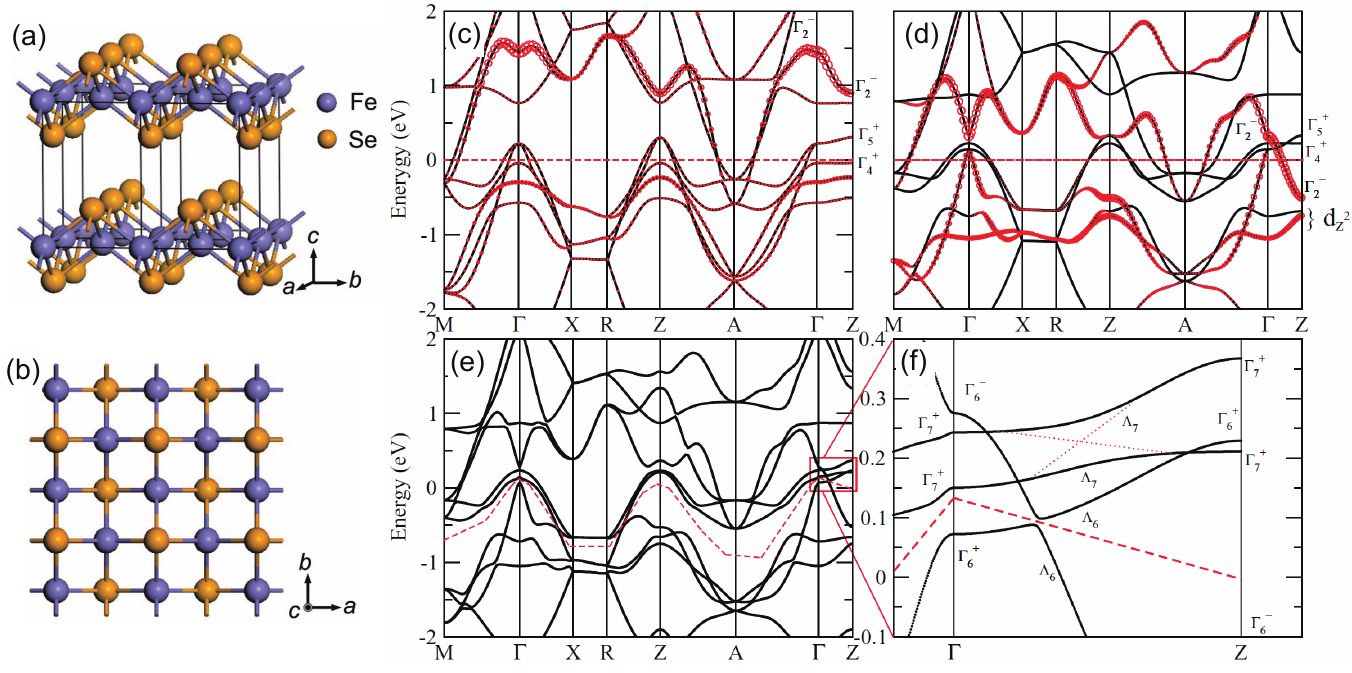}
    \caption{(a, b) Schematics of crystal structure of iron-chalcogenide, adopted from ref. \cite{Hsu_PNAS_2008} ($\copyright$2008 National Academy of Sciences, U. S. A.). (c, d) First-principles calculation results of band structure without the SOC for FeSe and for FeSe$_{0.5}$Te$_{0.5}$, respectively. The size of the red circles represent the contribution of the chalcogen $p_z$ orbitals. (e, f) Calculated band structure for FeSe$_{0.5}$Te$_{0.5}$ with the SOC. Figures of (c)-(f) are adopted from ref \cite{Wang_PRB_2015}. ($\copyright$2015 American Physical Society).}
    \label{fig_sim}
\end{figure}
\begin{figure}[h]
    \centering\includegraphics[width=14cm]{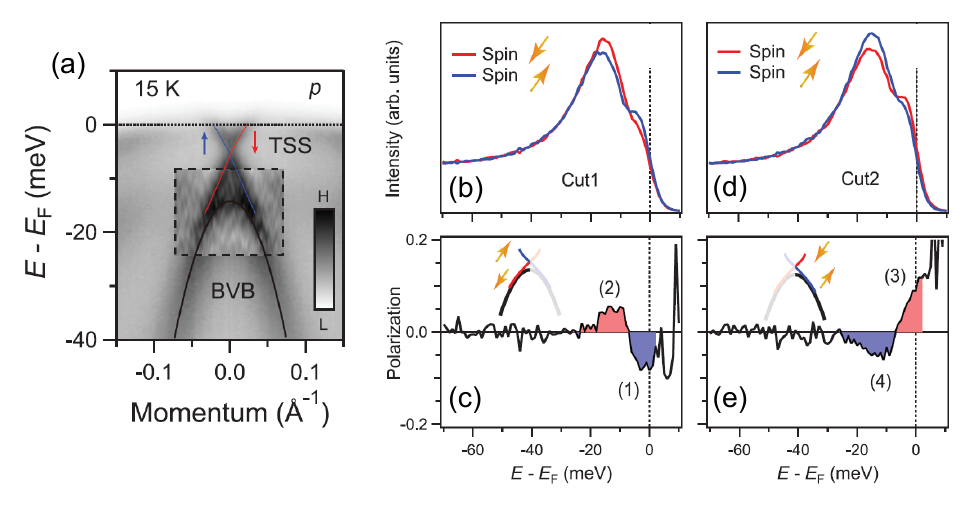}
    \caption{(a) Band dispersion in FeSe$_{0.45}$Te$_{0.55}$ along the $\Gamma-M$ direction recorded by ARPES. The spin-polarized Dirac surface states are indicated by colored lines. 
    (b, c) Spin-resolved energy distribution curves and their difference, respectively, taken on one side of the Dirac cone. 
    Spin polarizations are illustrated in the inset of (c). (d, e) Same as (b) and (c) but taken on the other side of the cone. Spin polarizations are reversed, consistent with the helical spin structure. All figures are adopted from ref \cite{Zhang_Science_2018}. ($\copyright$2018 The American Association for the Advancement of Science).}
\end{figure}

Key ingredients to make the Fe(Se,Te) topologically nontrivial are the strong spin-orbit coupling (SOC) and the interlayer coupling between the chalcogen $p_z$-orbitals~\cite{Wang_PRB_2015}.
According to the first principles calculation, the parent material FeSe has a topologically trivial band structure in which the odd-parity Se $p_z$ band is less dispersive and located far above even-parity iron $d$-bands along the $\Gamma-Z$ direction [Fig. 2(c)]. 
Two effects occur when $\gtrsim 50\%$ of Se is replaced with Te.
First, Te substitution enhances the interlayer coupling of the chalcogen $p_z$ orbitals since the Te $p_z$-orbital is more extended than that of Se.
This makes the $p_z$ band more dispersive and pushes it toward the Fermi level, giving rise to the band crossing between the $p_z$ and the iron $d$-bands ($d_{xz}, d_{yz}, d_{xy}$) along the $\Gamma-Z$ direction [Fig. 2(d)].
As a result, the band inversion occurs at the Z point.
Second, since Te is heavier than Se, one can expect a stronger SOC that opens a topologically nontrivial gap at one of the crossing points.
As a result, Fe(Se,Te) possesses a topologically nontrivial band structure, which naturally leads to the formation of a spin-polarized Dirac surface state.

The spin-polarized Dirac surface state was experimentally verified in FeSe$_{0.45}$Te$_{0.55}$ by using high-resolution spin-resolved ARPES (Fig. 3)~\cite{Zhang_Science_2018}.
Moreover, the measurements below $T_{\rm{c}}$ revealed that an isotropic superconducting gap opens on the observed spin-polarized surface states.
These observations suggest that the effective chiral $p$-wave superconductivity is realized at the surface.

Besides the topological nature, large energy separation $\delta E$ between the vortex bound states is expected in Fe(Se,Te) because it is in the crossover regime between the Bardeen-Cooper-Schrieffer superconductivity and Bose-Einstein condensation where $\epsilon_{\rm{F}}$ is comparable to $\Delta$.
According to ARPES and STM results, $\epsilon_{\rm{F}}$ is estimated to be $5 \sim 20$~meV~\cite{Rinott_SAdv_2017}, and the smallest $\Delta$ is about 1.5~meV~\cite{Hanaguri_Science_2010}.
Therefore, the energy of the lowest trivial CdGM state $E_{\mu=1/2} \sim \Delta^{2}/2\epsilon_{\rm{F}} = 60 \sim 220$~$\mu$eV.
This is two orders of magnitude larger than those in usual superconductors, providing an opportunity to resolve the discrete vortex bound states individually.
Nevertheless, the energy resolution of STM/STS must be better than a few tens of $\mu$eV to clearly distinguish the MZM from the lowest CdGM states that may appear as low as $60~\mu$eV.
Such a high energy resolution is only possible at ultra-low temperatures below 100~mK achieved by a dilution refrigerator.
The energy resolution of $20 \sim 30$~$\mu$eV has been achieved by a recently developed dilution-refrigerator-based STM/STS system, of which the attainable electron temperature is as low as 90~mK~\cite{Machida_RSI_2018}.

\section{Nature of the ZVBS in $\mathrm{Fe(Se,Te)}$}

\begin{figure}[h]
    \centering\includegraphics[width=15cm]{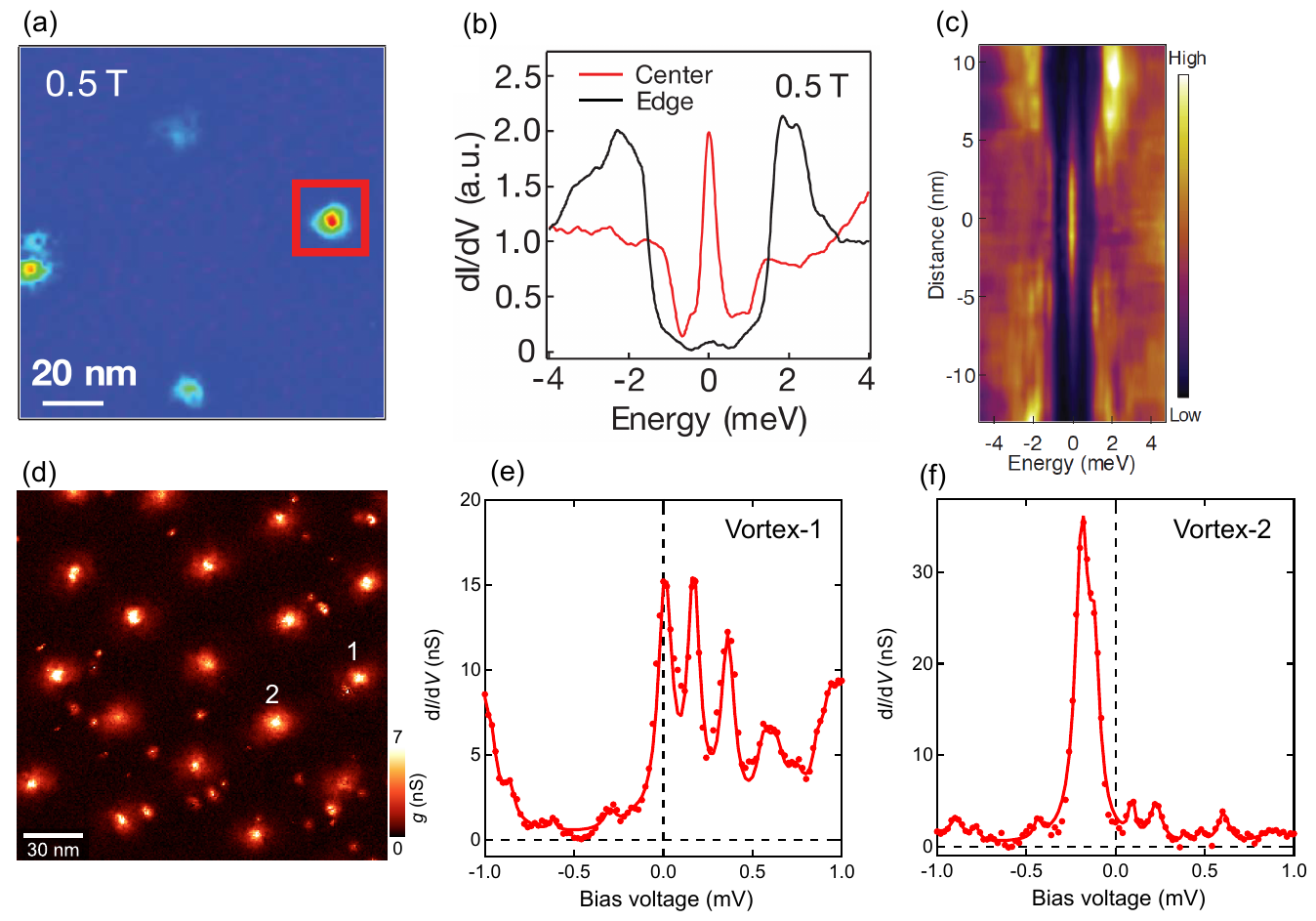}
    \caption{(a-c), Spectroscopic results of the vortex bound states of FeSe$_{0.45}$Te$_{0.55}$ taken at $B~=~0.5$~T using $^3$He-based STM with the energy resolution of $\sim~250~\mu$eV. (a) A zero bias conductance map. (b) tunneling spectra at the center (red) and away from (black) the vortex core. (c)  A color-intensity plot of spatial evolution of tunneling spectra around a vortex core. Figures of (a)-(c) are adopted from ref. \cite{Wang_Science_2018} ($\copyright$2018 The American Association for the Advancement of Science). (d-f), Spectroscopic results of the vortex bound states of FeSe$_{0.4}$Te$_{0.6}$ taken at $B~=~1$~T using DR-based STM with the energy resolution of $\sim~20~\mu$eV. (d) A zero bias conductance map. (e, f), High-energy resolution spectra taken at the core of vortices labeled as 1 and 2 in (d), respectively. There is a ZVBS in (e), but is not in (f). 
    }
\end{figure}

The STM/STS experiment on the putative MZM in the vortex cores of Fe(Se,Te) was first performed by Wang \textit{et al.}~\cite{Wang_Science_2018}.
They revealed an apparent ZVBS at the vortex core and claimed its MZM origin [Fig. 4(a)-(c)].
However, a subsequent experiment performed under similar conditions observed that the bound-state peaks appear only at finite energies, suggesting their trivial nature~\cite{Chen_NC_2018}.
Since the superconducting gap of Fe(Se,Te) is not spatially uniform~\cite{Massee_SAdv_2015}, the contradiction between the two experiments may be related to some kind of inhomogeneities.
Unfortunately, a recipe for preparing a uniform Fe(Se,Te) sample has not been established.
Besides the inhomogeneity problem, the above pioneering experiments were performed using $^{3}$He-refrigerator-based STM/STS systems with an energy resolution of $\sim 250$~$\mu$eV, which is insufficient to resolve the individual vortex bound states.
Therefore, the observed peaks in the spectrum may consist of multiple bound states.
We emphasize here that it is easy for the ZVBS to be misinterpreted as the finite-energy peak and for the finite-energy peak to be misinterpreted as the ZVBS, if the energy resolution of the STM/STS is worse than $\delta E$.

As mentioned above, the energy-resolution issue can be solved by adopting the dilution-refrigerator-based STM/STS system with an energy resolution of $20 \sim 30$~$\mu$eV.
Unfortunately, the inhomogeneity problem is currently unavoidable.
However, if one can inspect a large number of vortices at different locations and perform statistical analyses, it may help to identify the parameters essential for the ZVBS.
Such a high-energy-resolution experiment has been conducted on FeSe$_{0.4}$Te$_{0.6}$ where hundreds of vortices in different magnetic fields have been systematically investigated.
High-energy-resolution STM/STS found vortices with a ZVBS below the lowest possible energy ($\sim 60$~$\mu$eV) of the trivial CdGM state [Fig. 4(e)].
However, there are vortices without a ZVBS as well [Fig. 4(f)], indicating that some sort of inhomogeneity matters.

The spatial correlations between the presence or absence of the ZVBS and various preexisting quenched disorders, such as defects, local Te/Se ratio, and superconducting-gap size, were thoroughly investigated.
However, no discernible correlations were detected~\cite{Machida_NM_2019}.
Instead, it was found that the fraction of vortices with the ZVBS systematically decreases with increasing applied magnetic field [Fig. 5].
This suggests that inter-vortex distance may play a role in the ZVBS.
If there are two vortices with a MZM in each core and if these vortices come closer, the interaction between the trapped MZMs should be taken into account.
In such a case, the overlap of the Majorana wavefunctions causes the splitting of the MZM at zero energy into states at finite energies\cite{Cheng_PRL_2009}, resulting in an apparent disappearance of the ZVBS.
When vortices form a regular vortex lattice, many interacting MZMs form so-called Majorana bands at finite energies~\cite{Biswas_PRL_2013,Liu_PRB_2015,Chiu_PRB_2015,Affleck_PRB_2017}.
Although this interacting MZMs scenario seems to account for the observed field-induced disappearance of the ZVBS, the observed coexistence of vortices with and without the ZVBS remains elusive because all the vortices are identical in a regular lattice.
Apparently, the effect of inhomogeneity should be considered.
Given the observation that preexisting quenched disorders do not affect the ZVBS, it is reasonable to assume that the disorder in the vortex lattice plays an important role.
Indeed, vortices shown in Fig. 5(a) to (f) do not form regular lattices.
Large-scale theoretical simulations of the disordered vortex lattices in a 2-dimensional chiral $p$-wave superconductor reasonably reproduce the experimental observations~\cite{Chiu_SAdv_2020}.
\begin{figure}[h]
    \centering\includegraphics[width=15.5cm]{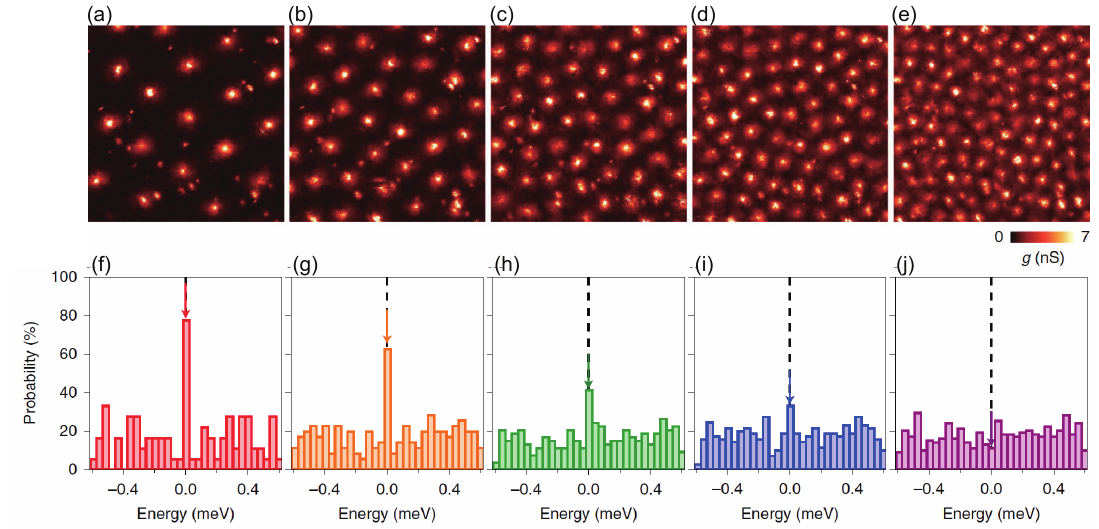}
    \caption{(a-e) Zero bias conductance maps in the same FOV at 1 T (a), 2 T (b), 3 T (c), 4 T (d) and 6 T (e).
    (f-j), The histograms of the appearance probability of the peaks in the tunnelling spectra at given energies taken at 1 T (f), 2 T (g), 3 T (h), 4 T (i) and 6 T (j).
    All the imaged vortices in this FOV were used for the analyses at all the $B$ values.
    The value at the zero energy of these histograms represent the fraction of vortices with the ZVBS at each magnetic field. }
    \label{fig_sim}
\end{figure}

\section{Search for qualitative Majorana features}

\subsection{Level sequence and conductance plateau}
The ZVBS observed in Fe(Se,Te) cannot be explained in terms of the topologically trivial CdGM state, suggesting its MZM origin.
However, the ZVBS is merely one of the necessary conditions for the MZM in the vortex core.
More direct and qualitative signatures expected for the MZM should be pursued.

Kong \textit{et al.} pioneered such an experiment~\cite{Kong_NP_2019}.
They focused on the level sequence of the vortex bound states $E_{\mu} = \mu(\Delta^2/\epsilon_{\rm{F}})$. 
As described above, the quantum number $\mu$ is a half-odd-integer ($\mu=\pm1/2,\pm3/2, \pm5/2, \dots$) for the $s$-wave superconductor~\cite{CAROLI1964307}, whereas $\mu$ is an integer ($\mu=0,\pm1, \pm2, \dots$) for the chiral $p$-wave superconductor where the MZM appears as the $\mu=0$ state at zero energy~\cite{Volovik_JETP_1999}.
This stark contrast in $\mu$ can be used to distinguish the vortex with the MZM from the trivial vortex.

Kong \textit{et al.} examined vortices with and without the ZVBS and showed that the level sequences roughly follow the integer and half-odd-integer level sequences, respectively ~\cite{Kong_NP_2019}.
They claim that sample inhomogeneities result in the phase separation of topological and non-topological regions, causing the presence or absence of the ZVBS~\cite{Kong_NP_2019}.
However, there still remain concerns.
First, as mentioned above, statistical analyses of the correlations between the quenched disorders and the presence or absence of the ZVBS exclude the possibility that the sample inhomogeneities themselves play a role~\cite{Machida_NM_2019}.
Second, the experiment by Kong \textit{et al.} was performed using a $^{3}$He-refrigerator-based STM with the energy resolution not enough to identify individual vortex bound states.
Therefore, there remain ambiguities in the determination of $\mu$, especially for the levels with larger $\mu$ that exceed the apparent superconducting-gap energy of $\sim 1.5$~meV.

It is indispensable to perform the level-sequence analyses for a large number of vortices with higher energy resolutions.
Figures 6(a) to (c) show the results of such analyses using the same data set for Fig.~5.
As shown in Fig. 6(a), there are multiple low-lying states even below the lowest-energy state in the previous work~\cite{Kong_NP_2019}.
We have indexed them from low to high energy with the ZVBS to be $n=0$ if it exists.
We then normalized the energy of the $n$-th peak $E_n$ by the energy of the lowest finite energy peak $E_1$ for more than 400 vortices to make histograms of distribution probabilities shown in Fig.~6(b) and (c).
For the integer and half-odd integer sequences, the peaks should appear at $|E_n/E_1| = n$ (solid line) and $|E_n/E_1| = 2n-1$ (dashed line), respectively.
The histograms are broad, with little differences between the two types of vortices with and without the ZVBS.
Apparently, the level-sequence analyses do not work properly for high-energy-resolution data.

There may be a couple of reasons that make the level-sequence analysis challenging.
First, the vortex bound states originating from the bulk bands may also be detectable at the surface.
Since Fe(Se,Te) is a multi-band superconductor, a multiple series of vortex bound states with different $\delta E = \Delta^2/\epsilon_{\rm{F}}$ can appear, making the simple indexing from low to high energy difficult.
Moreover, the tunneling spectrum without a magnetic field [Fig.~6(a)] shows many coherence-peak-like structures near the gap edge, which are apparently more than the number of bands crossing the Fermi level.
This means that, in addition to the multi-band effect, there is another effect that generates additional peaks in the tunneling spectrum.
Although the details are currently unknown, spatially inhomogeneous spectra may mean that the unavoidable sample disorders matter.
It is desirable to develop more uniform Fe(Se,Te) samples.
Nevertheless, we note that the simple disorder effect cannot explain the observed ZVBS, because any bound state should appear at finite energies in a trivial superconductor, except for an accidental situation.
\begin{figure}[h]
    \centering\includegraphics[width=14cm]{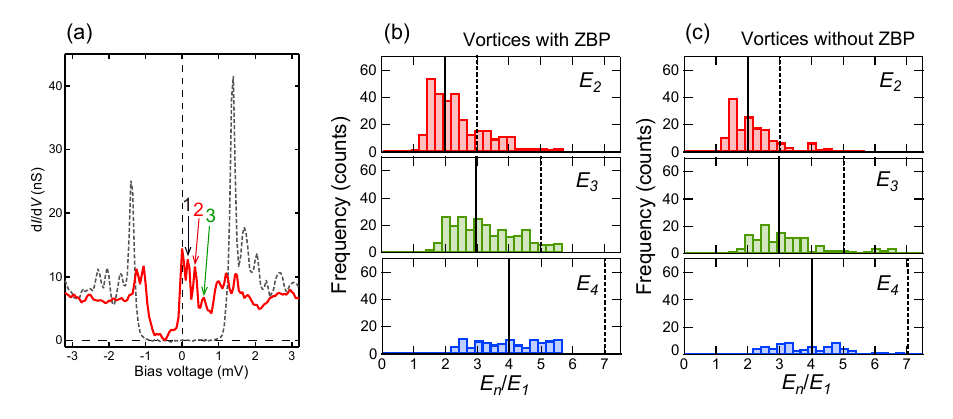}
    \caption{Results of the level sequence analysis using the DR-STM with the energy resolution of $\sim~20~\mu$eV. (a) Tunneling spectra taken at a vortex core with the ZVBS (red) and under zero magnetic field (gray dashed). (b, c), Histograms of the energies of the second (top), third (middle), and forth (bottom) lowest energy peaks for the vortices with and without the ZVBS, respectively.    
    The histograms of the energy of the $n$-th lowest finite energy peak normalized by the first one ($E_{n}$/$E_{1}$).
    These histograms are made from the tunneling spectra taken at $\sim$ 400 vortices with the energy resolution of $\sim$ 20 $\mu$eV. 
    Sold and dashed lines in each panel represent the expected energies for integer and half-odd integer sequence, respectively.
    }
    \label{fig_sim}
\end{figure}

Another important attempt to detect a feature unique to the MZM is the search for the quantized plateau of the tunneling conductance~\cite{Law_PRL_2009,He_PRL_2014,Haim_PRL_2015}.
Because of the perfect Andreev reflection associated with the equal superposition of an electron and a hole, the tunneling conductance of the MZM should be quantized at $G_0 = 2e^2/h~=~77.48~\mu$S, irrespective of the tip-sample separation~\cite{Law_PRL_2009,He_PRL_2014,Haim_PRL_2015}.
In the actual experiment, however, the thermal broadening and finite lifetime broadening (quasiparticle poisoning) may matter.
In this situation, the differential conductance at zero energy (namely, the height of the ZVBS) should monotonically increase with decreasing tip-sample distance and may eventually show a plateau structure at a certain value less than $G_0$~\cite{Budich_PRB_2012,Colbert_PRB_2014,Hu_PRB_2016,Setiawan_PRB_2017,Nichele_PRL_2017,Kong_NSR_2018}.

Zhu \textit{et al.} have performed such experiments on 60 vortices of FeSe$_{0.45}$Te$_{0.55}$ and observed the plateau-like behavior on half of them [Fig. 7(a)]~\cite{Zhu_Science_2020}.
However, since the experiment used a $^{3}$He-refrigerator-based STM with the energy resolution of $\sim$250 $\mu$eV, the observed ZVBSs should contain contributions from the trivial vortex bound states.
The energy resolution issue can be solved if we use the dilution-refrigerator-based STM, but there remains another technical problem.
Practically, before approaching the tip toward the sample surface to observe the conductance plateau, one has to stabilize the tip above the surface by a feedback loop that keeps the tunneling current constant to be typically $\sim 1$~nA at a typical bias voltage of $\sim 1$~mV.
Since the putative conductance plateau may appear at a current 1 or 2 orders of magnitude larger, the tip-sample distance has to be reduced by $100 \sim 200$~pm for typical tunneling conditions with the effective work function of a few eV.
However, as long as the tip and sample are both reasonably clean, the tunneling condition should break down above $\sim 10$~nA, resulting in a situation where the tip and sample are in contact~\cite{Gimzewski_PRB_1987,Zhang_NL_2011,Kim_Sadv_2017}.
We found that this jump-in-contact phenomenon occurs in Fe(Se,Te) as shown in Fig.~7(c).
It is difficult to control the junction in such a contact regime because the conditions are sensitive to the actual shape of the tip apex that deforms in an uncontrollable manner after the contact.
Therefore, the tip-sample distance dependence of the tunneling spectrum and the height of the ZVBS do not show systematic changes in the contact regime as shown in Fig.~7(d) and (e).
For the putative conductance plateau, it is necessary to guarantee that the tunneling current flows only through the MZM even in the contact regime.
\begin{figure}[h]
    \centering\includegraphics[width=15cm]{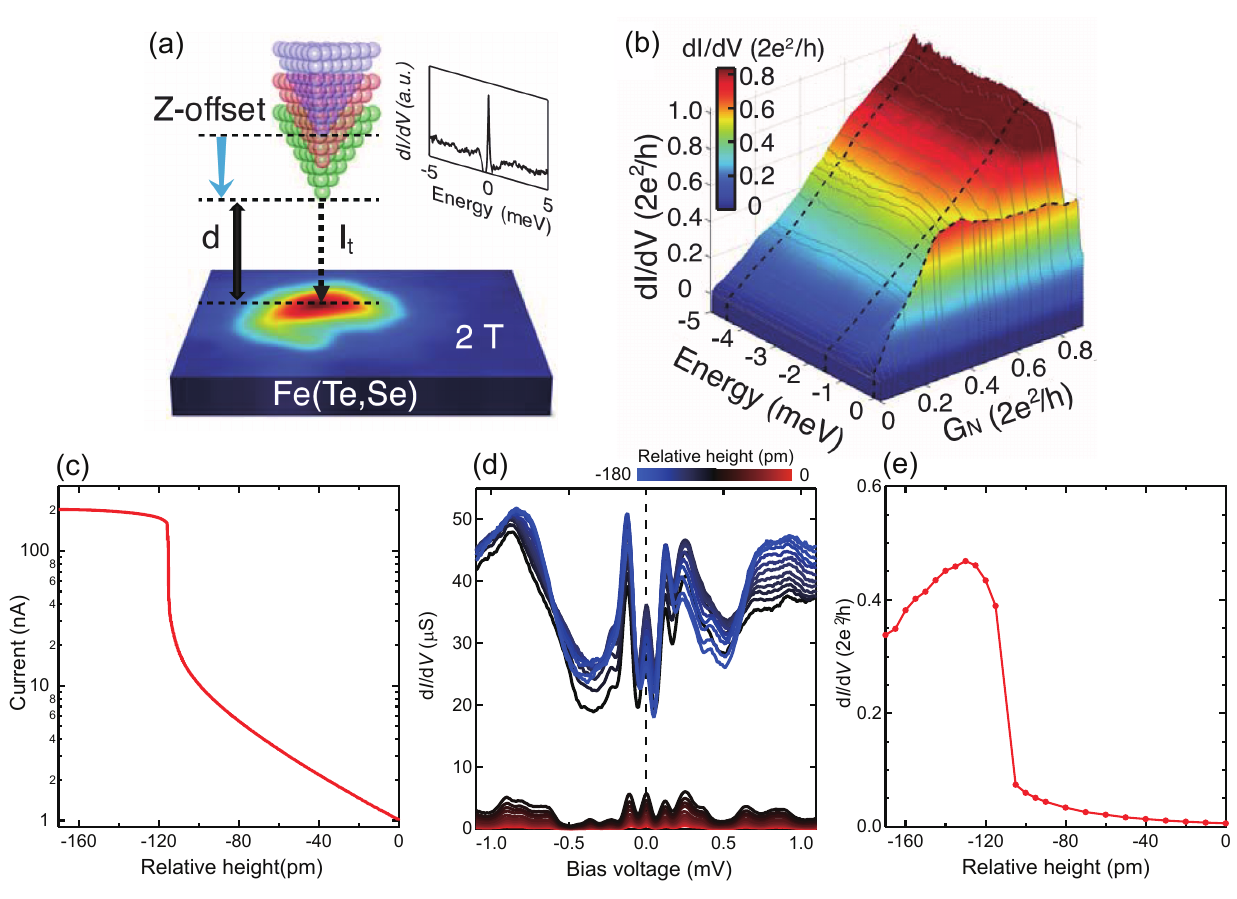}
    \caption{(a) A schematic illustration of the experimental setup for the variable-tunneling-coupling experiments. (b)Barrier height dependence of the tunneling spectra at a vortex core with the ZVBS. The zero bias peak indicates a plateau structure but others do not.
    Figures of (a) and (b) are adopted from ref. \cite{Zhu_Science_2020} ($\copyright$2018 The American Association for the Advancement of Science).
    (c-e) Tip-sample distance $Z$ dependence of tunneling current at -5 mV in (c), tunneling spectra in (d), and zero bias conductance in (e) taken at the vortex core of Fe(Se,Te) with the energy resolution of $\sim$ 20 $\mu$eV.}
\end{figure}

\subsection{Other potential experiments to detect the Majorananess}
Besides the level sequence and the quantized conductance plateau, more experiments have been considered to detect the "Majorananess".
One of the ideas is to detect the unique shot-noise expected in the current associated with the MZM~\cite{Bolech_PRL_2007,Nilsso_PRL_2008,Golub_PRB_2011}.
In general, the power of the shot-noise of the tunneling current $S$ is proportional to the effective charge per tunneling event $q_{\rm{eff}}$: $S = 2q_{\rm{eff}}|I|$. 
Because of the complete Andreev reflection, the effective charge per tunneling event for the junction with the MZM should be $q_{\rm{eff}} = 2e$, whereas $q_{\rm{eff}} = e$ in the case of the conventional single electron tunneling process.
This stark contrast can be used to distinguish the MZM from the trivial bound states~\cite{Bolech_PRL_2007,Nilsso_PRL_2008,Golub_PRB_2011}.
The shot-noise spectroscopy has been performed at the vortex core of Fe(Se,Te) using the recently-developed shot-noise STM~\cite{Bastiaans_RSI_2018} with a superconducting STM tip.
Although the observed $q_{\rm{eff}}$ was nearly $e$ even at the vortex core with the ZVBS, it is still unclear whether the ZVBS is associated with the MZM or not because the energy resolution was limited to be $\sim 250$~$\mu$eV~\cite{Ge_Arxiv_2022}.
Experiments at ultra-low temperatures with a higher energy resolution are anticipated.

Another possible experiment is the detection of the spin polarization expected in the vortex core with the MZM~\cite{Nagai_JPSJ_2014,Kawakami_PRL_2015}.
Since no spin polarization is expected in the trivial vortex core, a spin-sensitive experiment can provide an important clue.
In principle, STM can be spin-sensitive if a magnetic metal is used for the scanning tip~\cite{Wiesendanger_RSI_2009,Oka_RSI_2014}.
Since the principle of such a spin-polarized STM is the tunneling magneto-resistance effect, the larger the spin polarization of the tip, the better spin sensitivity is expected.
The spin polarization of a conventional magnetic tip is limited to a few tens of \%, and it has been difficult to estimate the absolute value of the spin polarization on the sample side.
Recently, a technique to achieve a 100\% spin-polarized tip has been developed, enabling us high-sensitivity and quantitative spin polarization measurements~\cite{Schneider_SAdv_2021,Huang_PRR_2021,Machida_PRR_2022}.
The idea is to use the Yu-Shiba-Rusinov (YSR) state, which is a 100\% spin-polarized bound state created near a magnetic impurity in a superconductor.
If a single magnetic atom is attached at the apex of the superconducting STM tip, the YSR state can act as a perfect spin filter for the tunneling current~\cite{Schneider_SAdv_2021,Huang_PRR_2021,Machida_PRR_2022}. 
Another important feature is that the narrowness of the YSR state in the spectrum is limited only by temperature~\cite{Machida_PRR_2022}, enabling us sub-meV energy resolution.
Therefore, spin-polarized spectroscopy using the YSR tip can be a promising tool to clarify the spin polarization of the ZVBS in Fe(Se,Te).

Besides the measurement techniques, more appropriate platforms that may host the MZM should be developed.
Although Fe(Se,Te) has great potential as the MZM platform because it naturally realizes the Fu-Kane proposal~\cite{Fu_PRL_2008} and has a large energy spacing between the vortex bound states, its chemical and electronic disorders make straightforward data interpretation difficult.
Although the source of the disorder is unclear, the substitution of Te for Se may cause inhomogeneities in the Fe-chalcogen layer.
Several other iron-based superconductors with stoichiometric Fe-chalcogen or Fe-arsenic layer have been proposed to host the chiral $p$-wave superconductivity at the surface.
These include (Li$_{1-x}$Fe$_x$)OHFeSe~\cite{Liu_PRX_2018} and CaKFe$_4$As$_4$~\cite{Liu_NC_2020}.
These compounds possess a $T_{\rm{c}}$ more than twice as high as that of Fe(Se,Te), resulting in a larger energy separation of vortex bound states.
Indeed, even by using $^3$He-refrigerator STM/STS systems, the ZVBSs were successfully resolved from the lowest trivial bound states in these materials~\cite{Liu_PRX_2018,Liu_NC_2020}.
Although these materials are more homogeneous than Fe(Se,Te), the vortex core spectra still vary from vortex to vortex.
In addition, these materials possess multiple cleaving planes, resulting in various surfaces that may exhibit different properties even if the MZM should be topologically protected.
It is an important challenge to develop a platform without these uncertain factors.

\section{Summary}
We have overviewed the recent STM/STS experiments on the ZVBS at the vortex core of Fe(Se,Te) where the effective chiral $p$-wave superconductivity is expected to be realized at the surface.
The ZVBS appears well below the lowest possible trivial vortex bound state, suggesting its MZM origin.
The systematic surveys of a large number of vortices exhibit the coexistence of the vortices with and without the ZVBS and that the fraction of vortices with the ZVBS systematically decreases with increasing the magnetic field, suggesting the crucial role of the Majorana-Majorana interaction.
The ZVBS is merely one of the necessary conditions for the MZM. Detecting unambiguous qualitative signatures of MZM is highly anticipated. A couple of attempts have been made, including the observation of the characteristic level sequence of the vortex with MZM and the search for the quantized conductance plateau, but they demand further investigations.
The doubling of the shot-noise intensity and the spin structure of the MZM have been proposed as potential proof of the "Majorananess" and are waiting for the experimental tests.
Once the MZM in the vortex core is proven, the next step toward the topological quantum computation should be the real-space manipulation of the vortex MZM.

\section*{Acknowledgment}
The authors acknowledge the collaboration of T. Tamegai, Y. Sun, S. Pyon, S. Takeda, Y. Kohsaka, C. J. Butler, T. Sasagawa, C.-K. Chiu, Y. Huang, F.-C. Zhang, and Y. Nagai.
This work was partly supported by CREST project JPMJCR16F2 and PRESTO project JPMJPR19L8 from the Japan Science and Technology
Agency, Grants-in-Aid for Scientific Research (KAKENHI) (numbers 17H01141, 16H04024, 19H01843), a Grant-in-Aid for Young Scientists (KAKENHI) (number 19K14661), and Japan China Bilateral Joint Research Project by the Japan Society for the Promotion of Science (JSPS).

\bibliography{PTEP_Majorana_Refs_0710}
\end{document}